# A new model and its physics

## Tian De Cao[*]

The high temperature superconductivity in cuprate materials[1] has puzzled scientists over twenty years. We must find a new way to understand superconductivity. It is found the spin-charge correlation may dominate the superconductivity[2], and we base our judgment upon the features of various superconductors. Thus we presented the idea that superconductivity could be described by correlations. To develop this idea into a quantitative theory, the first work is to give a model and show that various superconductivities can be included in this model. Moreover, superconductivity can originate from the spin-singlet pairing[3] or from the spin-triplet pairing[4]. The spin-singlet pairing favors to appear at the border of antiferromagnetism[5], while the spin-triplet pairing favors to appear at the border of ferromagnetism[6]. The coexistence between superconductivity and magnetism is also possible[7,8]. Therefore, the second work is to reveal the relation between superconductivity and magnetism.

Let us first find out a new model on the basis of the correlation superconducting mechanism. The correlations resulted from spins include the spin-charge correlation and the spin-spin correlation. In the general, the correlations lead to the local interaction

$$V^{(1)}_{\rho_1\rho_2}(x_1,x_2) = \int \lambda_{\rho_1 s'}(x_1,x')dx' \lambda_{s's''}(x',x'')dx'' \lambda_{s''\rho_2}(x'',x_2) \quad (1)$$

where $\lambda_{\rho s}$ and $\lambda_{ss}$ represent the effects of charge-spin correlation and the one of spin-spin correlation respectively. We have introduced electron coordinate $x=(\vec{x},s_z)$ with $s_z = \sigma \frac{1}{2}\hbar$, where the spin index $\sigma = \pm 1$. The integration over $x'$ and $x''$ is related to the propagator of spin excitation (imaginary) from $x'$ to $x''$. The time development is not shown in models for the Schrodinger representation. The form such as $\hat{V}_{\rho_1\rho_2}(x_1,x_2) = \int \kappa_{\rho_1 s'}(x_1,x')dx' \kappa_{s'\rho_2}(x',x_2)$ is incorrect, since interaction requires mediators, while this expression can not express the propagations of spin excitations. In next step, it is reasonable for one to assume $\lambda_{\rho s}(x,x') = \lambda^{(0)}_{\rho s}\delta(x-x') + \lambda^{(1)}_{\rho s}(x,x')$. The first term containing $\lambda^{(0)}_{\rho s}$ describes the effects of short length correlation, while $\lambda^{(1)}_{\rho s}$ describes the effects of long length correlation. In addition, because the magnetic susceptibility increases with the spin-spin correlation, $\lambda_{s_1 s_2} \propto \sigma_1 \cdot \sigma_2 \chi_m$, we can write it as $\lambda_{ss} = \gamma \sigma_1 \cdot \sigma_2 \chi_m$. The interaction associated with spin freedom is thus taken as the form $V^{(1)}_{\rho_1\rho_2}(x_1,x_2) = -\gamma \lambda^{(0)}_{\rho_1 s_1}\lambda^{(0)}_{s_2 \rho_2}\sigma_1 \cdot \sigma_2 \chi_m(\vec{x}_1-\vec{x}_2) + V'_{\rho_1\rho_2}(x_1,x_2)$. Because the spin-charge correlation and the charge-spin correlation have the same meaning, $\lambda^{(0)}_{s\rho} = \lambda^{(0)}_{\rho s}$, the interaction is also written as

$$V^{(1)}_{\rho_1\rho_2}(x_1,x_2)$$
$$= -q_1 q_2 g^2_{s\rho}\sigma_1 \cdot \sigma_2 \chi_m(\vec{x}_1-\vec{x}_2) + V'_{\rho_1\rho_2}(x_1,x_2) \quad (2)$$

where $g_{s\rho}$ is the constant number describing the effect of the spin-charge correlation, and the magnetic susceptibility $\chi_m$ describe the effect of spin-spin correlation. One can find that the first term in $V^{(1)}_{\rho_1\rho_2}$ is dominated by the short length interaction, while the second term in $V^{(1)}_{\rho_1\rho_2}$ is dominated by the long length interaction. Because the correlations associated with spins are dominated by the short length part, $V'_{\rho_1\rho_2}$ should be neglected. Similarly, the charge-charge correlation will lead

$$V^{(2)}_{\rho_1\rho_2}(x_1,x_2)$$
$$= \int \lambda_{\rho_1 \rho'}(x_1,x')dx' \lambda_{\rho'\rho''}(x',x'')dx'' \lambda_{\rho''\rho_2}(x'',x_2)$$

with the propagators of charge excitations from $x'$ to $x''$.

* *Department of physics, Nanjing University of Information Science & Technology, Nanjing 210044, China*

We can introduce the charge susceptibility $\chi_\rho$ and write

$$V^{(2)}_{\rho_1\rho_2}(x_1,x_2) = -q_1 q_2 g^2_{\rho\rho}\chi_\rho(\vec{x}_1-\vec{x}_2) + V''_{\rho_1\rho_2}(x_1,x_2)$$

The charge susceptibility $\chi_\rho$ is not the electronic susceptibility $\chi_e$, here $\chi_\rho$ and $\chi_m$ have the same physical unit as soon as both $g_{\rho\rho}$ and $g_{\rho s}$ take the same unit.

On the basis of the consideration above, the total (affective) interaction between quasiparticles is thus written as

$$V_{\rho\rho}(x_1,x_2) = -q_1 q_2 g^2_{\rho\rho}\chi_n(\vec{x}_1-\vec{x}_2) + V^{other}_{\rho\rho}(x_1,x_2)$$
$$-q_1 q_2 g^2_{s\rho}\sigma_1\cdot\sigma_2\chi_m(\vec{x}_1-\vec{x}_2) \quad (3)$$

where $V^{other}_{\rho\rho} = V'_{\rho_1\rho_2} + V''_{\rho_1\rho_2}$. The first term in (3) describe the effect of charge-charge correlation in which charge excitations (such as phonons) are mediators, while the third term contains the effects of both spin-charge correlation and spin-spin correlation in which spin excitations are mediators. Although Anderson argued that attraction between particles could not appear in a model with actual parameters[9], it is conceived that the affective interaction between particles can have attraction regions in space for an anisotropic material. $V^{other}_{\rho\rho}$ is to be neglected for strongly correlated systems provided the long length effects is not dominant. Thus we get the affective interaction

$$V_{\rho\rho}(x_1,x_2) = -q_1 q_2 g^2_{\rho\rho}\chi_n(\vec{x}_1-\vec{x}_2)$$
$$-q_1 q_2 g^2_{s\rho}\sigma_1\cdot\sigma_2\chi_m(\vec{x}_1-\vec{x}_2) \quad (4)$$

This expression is similar to the interaction of Monthoux's expression[10], but some difference exists between them. It seems that the first and the second term in (4) could mediate superconductivity separately. However, on our evaluation, if the second term in (4) could be neglected, the first and the second term in (3) should give a repulsive interaction. Therefore, the second term in (4) can not be neglected in form. BCS theory[11] suggested superconductivity is induced by phonons, while we suggest superconductivity is induced by all excitations measured with spin-charge correlation.

Therefore, the effects of charge-charge correlation on superconductivity in (4) can be attributed into the effect of spin-charge correlation. Because $g_{s\rho}\neq 0$ for a superconductor as discussed above, we can introduce the affective susceptibility $\chi$ to sum (4) to the affective interaction

$$V_{\rho\rho}(x_1,x_2) = -q_1 q_2 g^2_{s\rho}\sigma_1\cdot\sigma_2\chi(\vec{x}_1-\vec{x}_2) \quad (5)$$

The affective susceptibility $\chi$ is not, but is near, the magnetic susceptibility $\chi_m$.

A popular basis set is taken as plane waves, and this leads the second quantization of the interaction to this form

$$H_{int} = -e^2 g^2_{s\rho}\sum_{\substack{k,k',q\\\sigma,\sigma'}}\sigma\sigma'\chi(q)c^+_{k+q,\sigma}c_{k\sigma}c^+_{k'-q,\sigma'}c_{k'\sigma'} \quad (6)$$

thus the affective Hamilton of particles is

$$H = \sum_{k,\sigma}\xi_k c^+_{k\sigma}c_{k\sigma}$$
$$-e^2 g^2_{s\rho}\sum_{\substack{k,k',q\\\sigma,\sigma'}}\sigma\sigma'\chi(q)c^+_{k+q,\sigma}c_{k\sigma}c^+_{k'-q,\sigma'}c_{k'\sigma'} \quad (7)$$

This model is usually easy to be solved if only the susceptibility function $\chi$ is given. This means that both strongly- and weakly- correlated electrons can be described with a weak-correlation model. The lattice Fourier transforms lead (7) to the t-$\chi$ model

$$H = \sum_{l,l',\sigma}t_{ll'}c^+_{l\sigma}c_{l'\sigma} - 4e^2 g^2_{s\rho}\sum_{l,l'}\chi(l-l')\hat{S}_{lz}\hat{S}_{l'z} \quad (8)$$

where we have taken the chemical potential $\mu=0$. If we intend to determine the direction of spins, we must extend the model to the case of $\hat{S}_z\to\hat{\vec{S}}$. The model (8) is similar to the t-J model $H_{t-j} = -t\sum_{<l,l'>,\sigma}c^+_{l\sigma}c_{l'\sigma} + J\sum_{<l,l'>}\hat{\vec{S}}_l\cdot\hat{\vec{S}}_{l'}$.

We have taken the mark $l\equiv\vec{R}_l$, thus the position vector between electrons is $\vec{r} = \vec{R}_l - \vec{R}_{l'} \equiv l - l'$.

There are differences between these two models. The t-J model is based on the strongly correlated features of electrons, while the t-$\chi$ model is based on the correlation superconducting mechanism which is also applicable to the weak correlation systems. There are the same aspects between them, too. Since ferromagnetism or antiferromagnetism can be included in the well-known t-J model with negative or positive $J$, similarly, these magnetisms can also be included in the t-$\chi$ model.

Now let us see the ferromagnetism of this model in normal state. Define the Green function

$$G(k\sigma, \tau - \tau') = -<T_\tau c_{k\sigma}(\tau) c_{k\sigma}^+(\tau')> \qquad (9)$$

after establishing the dynamical equations of these functions in normal state, we get

$$\frac{\partial}{\partial \tau} G(k\sigma, \tau - \tau')$$
$$= -\delta(\tau - \tau') - \tilde{\xi}_{k\sigma} G(k, \sigma, \tau - \tau') \qquad (10)$$

where

$$\tilde{\xi}_{k\sigma} = \xi_k - 2e^2 g_{s\rho}^2 \sigma \chi(0) n_i$$
$$+ e^2 g_{s\rho}^2 \sum_q [\chi(q) + \chi(-q)] G(k+q\sigma, \tau=0) \qquad (11)$$

and the chemical potential is redefined. It is easy to find $G(k\sigma, i\omega_n) = \dfrac{1}{i\omega - \tilde{\xi}_{k\sigma}}$. This Green function leads to the possible spin component per each site

$$\bar{S}_{iz} = \frac{1}{2}[n_F(\tilde{\xi}_{k\uparrow}) - n_F(\tilde{\xi}_{k\downarrow})] \qquad (12)$$

where

$$n_\uparrow = \sum_k n_F(\tilde{\xi}_{k\uparrow}) \text{ and } n_\downarrow = \sum_k n_F(\tilde{\xi}_{k\downarrow}) \qquad (13)$$

Because it is usually $\tilde{\xi}_{k\uparrow} < \tilde{\xi}_{k\downarrow}$ for $\chi(0) > 0$ with the use of Eq. (11), along the line determined by Eq. (11)-(13), we find the solution of ferromagnetism $n_\uparrow - n_\downarrow > 0$. Similarly, consider the nearest interaction and divide the lattice into two sublattice, we find the antiferromagnetism can be also included in the model (7).

To consider the spin-singlet pairing, we define the correlation function

$$F^+(k\sigma, \tau - \tau') = <T_\tau c_{k\sigma}^+(\tau) c_{\bar{k}\bar{\sigma}}^+(\tau')> \qquad (14)$$

and establish the dynamical equation

$$\frac{\partial}{\partial \tau} F^+(k\sigma, \tau - \tau')$$
$$= \tilde{\xi}_{k\sigma} F^+(k\sigma, \tau - \tau')$$
$$- e^2 g_{s\rho}^2 \sum_q [\chi(q) + \chi(-q)] F^+(k+q\sigma, \tau=0)$$
$$\cdot G(\bar{k}\bar{\sigma}, \tau - \tau') \qquad (15)$$

For the solution $G(k\sigma) = G(\bar{k}\bar{\sigma}) \equiv G(k) = G(-k)$, we get $\tilde{\xi}_{k\sigma} \equiv \tilde{\xi}_k = \tilde{\xi}_{\bar{k}}$, and this is in accord with the case of crystal material without magnetism. Using the Fourier transformation, we get the equations with spin-singlet pairing

$$(-i\omega_n + \tilde{\xi}_k) G(k\sigma, i\omega_n) = -1$$
$$+ e^2 g_{s\rho}^2 \sum_q [\chi(q) + \chi(-q)] F(k+q\sigma, 0) \ F^+(\bar{k}\bar{\sigma}, i\omega_n) \qquad (16)$$

$$(-i\omega_n - \tilde{\xi}_k) F^+(k\sigma, i\omega_n) =$$
$$- e^2 g_{s\rho}^2 \sum_q [\chi(q) + \chi(-q)] F^+(k+q\sigma, 0) G(\bar{k}\bar{\sigma}, i\omega_n) \qquad (17)$$

If $F(k\sigma, 0) = F^+(\bar{k}\bar{\sigma}, 0)$ and $F(k\sigma, 0)$ is real, we get the functions

$$G(k\sigma, i\omega_n) = \frac{1}{i\omega - \tilde{\xi}_k - \Delta_{s,d}^2(k\sigma)/(i\omega + \tilde{\xi}_k)} \qquad (18)$$

and

$$F^+(k\sigma, i\omega_n) = \frac{\Delta_{s,d}(k\sigma)}{i\omega + \tilde{\xi}_k + \Delta_{s,d}^2(k\sigma)/(i\omega - \tilde{\xi}_k)} \qquad (19)$$

where

$$\Delta_{s,d}(k\sigma) = |e| g_{s\rho}^2 \sum_q [\chi(q) + \chi(-q)] F(k+q\sigma, 0)$$

Therefore, the gap equation is

$$\Delta_{s,d}(k\sigma)$$
$$= e^2 g_{s\rho}^2 \sum_q [\chi(q-k) + \chi(-q+k)] \Delta_{s,d}(q\sigma)$$
$$\cdot \frac{n_F(E_{q\sigma}) - n_F(-E_{q\sigma})}{2 E_{q\sigma}}$$

or the form

$$\Delta_{s,d}(k)$$

$$= e^2 g_{s\rho}^2 \sum_q [\chi(q-k)+\chi(-q+k)] \Delta_{s,d}(q)$$

$$\cdot \frac{n_F(E_q)-n_F(-E_q)}{2E_q} \qquad (20)$$

where $E_k = \sqrt{\tilde{\xi}_k^2 + \Delta_{s,d}^2(k)}$, all functions do not depend on the spin index. This equation is similar to the well-known BCS gap equation. If $\chi(q)$ and $\Delta_p(k)$ are approximately taken as the constant near the Fermi surface, we require $\chi(0) < 0$ for $T_c > 0$, this is the same as the requirement of antiferromagnetism.

For the spin-triplet pairing, we define the correlation function

$$F_p^+(k\sigma, \tau-\tau') = <T_\tau c_{k\sigma}^+(\tau) c_{\bar{k}\sigma}^+(\tau')> \qquad (21)$$

With the similar calculation above, we get the gap function

$$\Delta_p(k) = -|e| g_{s\rho}^2 \sum_q [\chi(q-k)+\chi(-q+k)] \Delta_p(q)$$

$$\cdot \frac{n_F(E_q)-n_F(-E_q)}{2E_q} \qquad (22)$$

where $\Delta_p(k) = |e| g_{s\rho}^2 \sum_q [\chi(q)+\chi(-q)] F_p(k+q\sigma, 0)$, and $E_k = \sqrt{\tilde{\xi}_k^2 + \Delta_p^2(k)}$. If $\chi(q)$ and $\Delta_p(k)$ are approximately taken as the constant near the Fermi surface, we require $\chi(0) > 0$, this is the same as the requirement of ferromagnetism found above.

In summary, a phenomenological model is obtained from the correlation superconducting mechanism, superconductivity and magnetism can be described with this t-$\chi$ model. At the end, we give the following explanation or conclusion:

(1) The vector symbols of wave vectors have been neglected in all expressions, $k \equiv \vec{k}$. The susceptibility $\chi(k)$ should be anisotropic for strongly correlated electrons, and this favors the d-wave pairing or the anisotropic s-wave pairing.
(2) They have a symbol difference in equations (20) and (22), thus the susceptibility functions should favor different forms for the spin-singlet pairing and the spin-triplet pairing separately.
(3) Whether the pairing is s- or d- wave symmetry depends on the susceptibility function, no matter what the pairing belongs to the spin-singlet or the spin-triplet.
(4) The interaction associated with the spin-triplet pairing allows the weak ferromagnetism or the short-length ferromagnetic order, while the interaction associated with the spin-triplet pairing allows the weak antiferromagnetism or the short-length antiferromagnetic order. This conclusion is concerned with the coexistence between superconductivity and magnetism, although this has to be proved with calculation.